\documentclass[fleqn,12pt,twoside]{article}
\usepackage{espcrc1}

\usepackage{graphicx}
\usepackage[figuresright]{rotating}

\newcommand{\chiPT}{$\chi$PT}

\newcommand{\AmS}{{\protect\the\textfont2
  A\kern-.1667em\lower.5ex\hbox{M}\kern-.125emS}}

\title{Observing Chiral Nonanalytic Behavior with FLIC Fermions}

\author{%
D. B. Leinweber\address[CSSM]{Centre for the Subatomic
               Structure of Matter and Department of Physics\\
               University of Adelaide, Adelaide SA 5005 Australia},
A. W. Thomas\addressmark,
A. G. Williams\addressmark,
R. D. Young\addressmark,
J. M. Zanotti\addressmark\address{John von Neumann-Institut f$\ddot{\rm u}$r
             Computing NIC, \\ 
             Deutches Elektronen-Synchrotron DESY, D-15738 Zeuthen, Germany}
and
J. B. Zhang\addressmark[CSSM]
        }
       
\begin{document}

\maketitle

\vspace{-7cm}
\null \hfill ADP-03-126/T562
\vspace{7cm}
\vspace{-24pt}

\begin{abstract}
First results from lattice QCD revealing the chiral nonanalytic
behavior of nucleon and $\Delta$ baryon magnetic moments are
presented.  Numerical simulations in the light quark mass regime
employing the nonperturbatively ${\mathcal O}(a)$-improved conserved
vector current are enabled via FLIC fermions.  Quenched chiral
perturbation theory for the nucleon and $\Delta$ magnetic moments is
derived to next to next to leading nonanalytic order.  Numerical
simulation results for the proton and $\Delta$ baryon magnetic moments
in quenched QCD reveal dramatic signatures of the quenched meson
cloud, which are in accord with the predictions of quenched chiral
perturbation theory.
\end{abstract}

\section{NUMERICAL SIMULATIONS WITH FLIC FERMIONS}

Access to leading-edge supercomputing resources coupled with advances
in the formulation of computationally-inexpensive chirally-improved
lattice fermion actions \cite{Zanotti:2001yb,Leinweber:2002bw} enable
the numerical calculation of hadron structure in the chiral regime.
In this regime, the pseudo-Goldstone boson dressings of hadrons give
rise to significant non-analytic curvature in the quark-mass
dependence of observables.  The magnetic moments of baryons have been
identified \cite{Leinweber:2001jc,Leinweber:2002qb} as providing an
excellent opportunity for the direct observation of chiral nonanalytic
behavior in lattice QCD, even in the quenched approximation.

The numerical simulations of the electromagnetic form factors
presented here are carried out using the Fat Link Irrelevant Clover
(FLIC) fermion action \cite{Zanotti:2001yb,Leinweber:2002bw} in which
the irrelevant operators introduced to remove fermion doublers and
lattice spacing artifacts are constructed with smoothed links.  These
links are created via APE smearing \cite{ape}; a process that averages
a link with its nearest transverse neighbors in a gauge invariant
manner.  Iteration of the averaging process generates a ``fat'' link.

The use of links in which short-distance fluctuations have been
removed simplifies the determination of the coefficients of the
improvement terms in both the action and its associated conserved
vector current.  Perturbative renormalizations are small for smeared
links and the mean-field improved coefficients used here are
sufficient to remove ${\mathcal O}(a)$ errors, in the lattice spacing
$a$, from the lattice fermion action.
The key is that both the energy dimension-five Wilson and Clover terms
\cite{Bilson-Thompson:2002jk} are constructed with smooth links, while
the relevant operators, surviving in the continuum limit, are
constructed with the original untouched links generated via standard
Monte Carlo techniques.

FLIC fermions provide a new form of nonperturbative ${\mathcal O}(a)$
improvement \cite{Leinweber:2002bw,inPrep} where near-continuum
results are obtained at finite lattice spacing.  Access to the light
quark mass regime is enabled by the improved chiral properties of the
lattice fermion action.  The magnitude of additive mass
renormalizations is suppressed \cite{inPrep} which otherwise can
lead to singular behavior in the propagators as the quarks become
light.

The ${\mathcal O}(a)$-improved conserved vector current
\cite{Martinelli:ny} is used.  Nonperturbative improvement is achieved
via the FLIC procedure where the terms of the Noether current having
their origin in the irrelevant operators of the fermion action are
constructed with mean-field improved APE smeared links.  The
preliminary results presented here are from a sample of 255 $20^3
\times 40$ mean-field improved Luscher-Weisz \cite{Luscher:1984xn}
gauge field configurations having a lattice spacing of 0.128 fm as
determined by the Sommer scale $r_0=0.50$ fm.

\section{CHIRAL NONANALYTIC BEHAVIOR}

The truncation of the low-energy expansion of chiral effective field
theory introduces errors into the predictions of chiral perturbation
theory (\chiPT).  In the process of a simple truncation, one sets the
coefficients of higher-order terms of the expansion (both analytic and
nonanalytic) to zero by hand.  While such a procedure is often
described as ``systematic'' or ``model independent'' the truth is that
the coefficients of these higher-order terms are generally not zero.
Thus, the truncated expansion is a poor representation of the chiral
expansion of QCD.  In this case, ``model independent'' simply means
that no attempt has been made to estimate the coefficients of higher
order terms in the expansion.  It is not necessarily a good feature.

While such an approach might be forgiven if the coefficients of the
higher order terms were indeed small, there is now mounting evidence
that this is not the case.  For the nucleon mass, the best
determination of the low-energy constants \cite{Young:2002ib} from the
physical nucleon mass and state-of-the-art lattice QCD results
\cite{AliKhan:2001tx} indicate the nucleon mass has the following
chiral expansion (in appropriate powers of GeV)
\begin{eqnarray}
m_N &=& (0.897\pm0.001)  \, + \, (2.84\pm0.04) \, m_\pi^2 \, + \, \chi_3
\, m_\pi^3 + (22.0\pm1.6) \, m_\pi^4 \, \nonumber \\
&&+ \, \chi_4 \, m_\pi^4 \log\left ( m_\pi^2/ 1\ {\rm GeV}^2 \right) \, 
+ \, \chi_5 \, m_\pi^5 \, + \, \cdots \, ,
\label{dimRegExp}
\end{eqnarray}
where $\chi_i$ are the known model-independent coefficients of the
leading nonanalytic terms of the expansion and the quoted
uncertainties are purely systematic \cite{Young:2002ib}.  The
coefficient of the $m_\pi^6$ term from the $N \to N \pi$ self energy
alone is $-75\pm35\ {\rm GeV}^{-5}$.  Hence the ``systematic''
approach of setting the coefficients of all higher order terms to zero
can be troublesome for the power-series like expansion of
dimensional-regularization (DR).

Fortunately there is a way to estimate the coefficients of the higher
order terms, while preserving the model-independent features of chiral
effective field theory to the chiral-counting order that one is
working.  Through the process of regulating loop integrals via a
finite-range regulator (FRR) \cite{Young:2002ib}, one re-sums the chiral
expansion in a manner which preserves the model-independent features
of chiral perturbation theory.  For example, the coefficients of the
leading nonanalytic terms of the expansion are preserved exactly.
However, in the process of expanding the expressions of FRR \chiPT\ to
recover the expansion of DR in Eq.\ (\ref{dimRegExp}), one also
encounters higher-order terms of the chiral expansion, whose
coefficients are functions of the regulator parameter ($\Lambda$)
governing the finite range of the regulator.  By optimizing $\Lambda$
in a fit of the FRR expansion to lattice QCD data, one obtains
estimates for the higher order terms of the chiral expansion, while
maintaining complete model-independence to the chiral order one is
working.

The introduction of a finite range regulator opens the question of the
functional form of the regulator.  Clearly one needs a functional form
that preserves the low energy physics of the chiral expansion, while
suppressing high-energy contributions where the internal structure of
the effective fields becomes important.  In practice, sharp-cutoff,
monopole, dipole and Gaussian vertex regulators have been investigated
in some detail.  Provided one allows the regulator parameter to be
constrained by lattice QCD data, a remarkable robustness is observed
in the predictions of the chiral expansion.  Systematic errors for the
nucleon mass have been estimated at less than 1\%, provided a smooth
FRR (monopole, dipole, Gaussian) is selected \cite{Young:2002ib}.
These regulators not only provide estimates for the coefficients of
higher-order analytic terms of the expansion, but also provide
estimates for the coefficients of non-analytic terms which must also
appear in the expansion.  This latter feature of the smooth regulators
has been identified as key to the success of the smooth regulators
over the sharp cutoff \cite{Young:2002ib}.


We use the diagrammatic method for evaluating the quenched chiral
coefficients of leading nonanalytic terms in heavy-baryon quenched
\chiPT\ \cite{Leinweber:2001jc,Leinweber:2002qb}.  Results for the
proton magnetic moment to next to next to leading nonanalytic (NNLNA)
order \cite{Leinweber:2002qb,Savage:2001dy} are generalized to the FRR
approach used here.  In quenched QCD, the $\Delta$ form factors are
simply proportional to the charge of the baryon
\cite{Leinweber:1995ie}.  Hence, consideration of the $\Delta^{++}$
charge state is sufficient to determine the chiral expansion for all
charge states.

\begin{figure}[tb]
\begin{center}
{\includegraphics[height=12cm,angle=90]{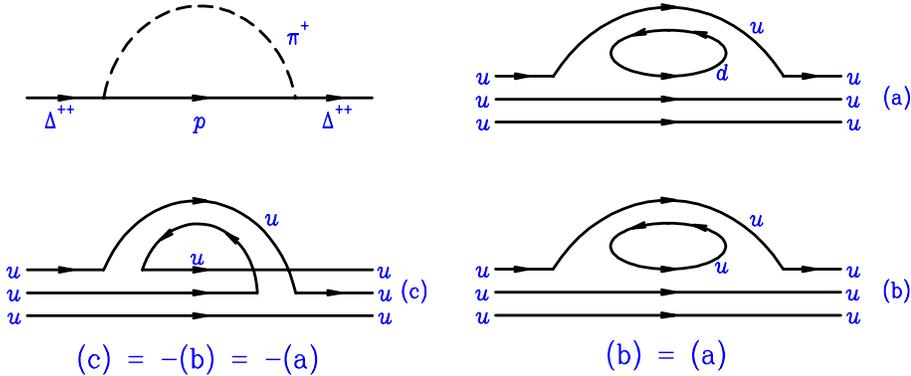}}
\vspace*{-0.8cm}
\caption{Quark-flow diagrams for meson-cloud contributions to 
  $\Delta^{++}$ in full QCD.} 
\label{deltaFlow}
\end{center}
\vspace{-1.2cm}
\end{figure}

The presence of the $\Delta \to N \pi$ decay channel is particularly
important for the mass dependence of $\Delta$ properties.  Rapid
curvature associated with nonanalytic behavior is shifted to larger
pion masses near the $N$-$\Delta$ mass splitting, $m_\pi \sim M_\Delta
- M_N$.  As described below, quenched-QCD decay-channel contributions
come with a sign opposite to that of full QCD.  This artifact holds
tremendous promise for revealing unmistakable signatures of the
quenched meson cloud.  

The change in sign for the decay-channel contributions is easily
understood through the consideration of the quark flow diagrams of
Fig.\ \ref{deltaFlow}, illustrating the meson-cloud contributions to
the $\Delta^{++}$ resonance in full QCD.  Quark flow diagram (a)
corresponds to the hadronic process described at left.  Since QCD is
flavor-blind, the process illustrated in diagram (b) is equivalent to
diagram (a) provided the masses of the $u$ and $d$ quarks are taken to
be equal.  On its own, diagram (b) describes the decay of the
$\Delta^{++}$ to a doubly-charged $uuu$ ``proton,'' which we denote
$p^{++}$.  Of course, such states do not exist in full QCD and diagram
(c) makes contributions exactly equal but opposite in sign to diagram
(b) when the intermediate state is a $uuu$ proton.  Upon quenching the
theory, both diagrams (a) and (b) are eliminated, leaving only diagram
(c).  Hence the physics of the $\Delta \to N \pi$ decay is present in
the quenched approximation \cite{Labrenz:1996jy} but its contribution
has the wrong sign.

\begin{figure}[tb]
\begin{center}
{\includegraphics[width=10cm,angle=0]{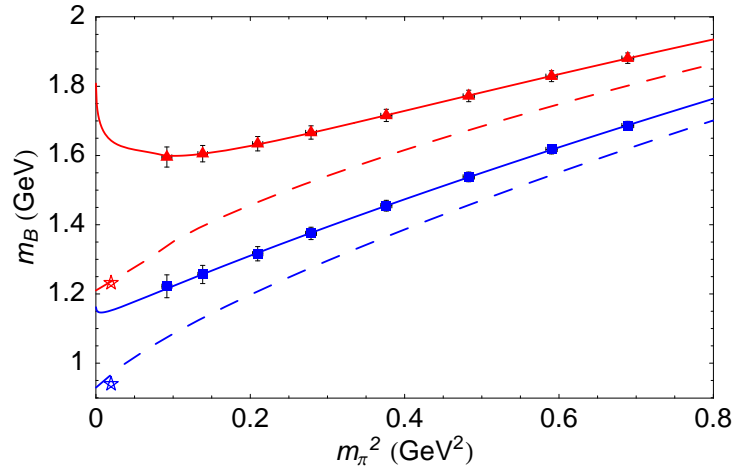}}
\vspace*{-1.0cm}
\caption{FLIC fermion simulation results for the masses of
  the proton ($\circ$) and $\Delta$ resonance ($\triangle$) in
  quenched QCD.  Solid curves indicate the fits of FRR quenched \chiPT\
  to the lattice simulation results, while dashed curves indicate the
  one-loop correction to the quenched approximation
  \protect\cite{Young:2002cj}.  Stars denote the physical values.}
\label{masses}
\end{center}
\vspace{-1.0cm}
\end{figure}

This aspect of the quenched $\Delta$ is the predominant feature giving
rise to the flattening of the $\Delta$ mass as a function of quark
mass at the lightest FLIC-fermion quark masses depicted in Fig.\
\ref{masses}.  Double-hairpin $\eta^\prime$ contributions impact the
$\Delta$ mass at much lighter quark masses most notably between the
chiral and physical pion mass.

As for the proton, the double-hairpin $\eta^\prime$ dressing
$\Delta^{++} \to \Delta^{++} \eta^\prime$ provides the LNA
contribution to the $\Delta^{++}$ magnetic moment, generating a
logarithmic divergence in the chiral limit.  For electromagnetic form
factors, Fig.\ \ref{deltaFlow}(c) indicates that the coefficient of
the NLNA contribution proportional to $m_\pi$ will vanish in the
quenched approximation because of the neutral charge of the meson.
However, Fig.\ \ref{deltaFlow}(c) will make significant contributions
when the electromagnetic current couples to the intermediate $p^{++}$.
We estimate the tree-level magnetic moment of the $uuu$ proton using
standard SU(6) symmetry, $\mu_{p^{++}}=\frac{4}{3}\mu_u -
\frac{1}{3}\mu_u = \frac{2}{3}\mu_p = \frac{1}{3}\mu_{\Delta^{++}}$.
The quark flow diagrams of Fig.\ \ref{deltaFlow} also include
contributions from $\Delta$ intermediate states.  In terms of the full
QCD process $\Delta^{++} \to \Delta^{++} \pi^0$ we find the total
quenched contribution to be $(4/3) \left ( \Delta^{++} \to \Delta^{++}
\pi^0 \right )$.

\begin{figure}[tb]
\begin{center}
{\includegraphics[width=10cm,angle=0]{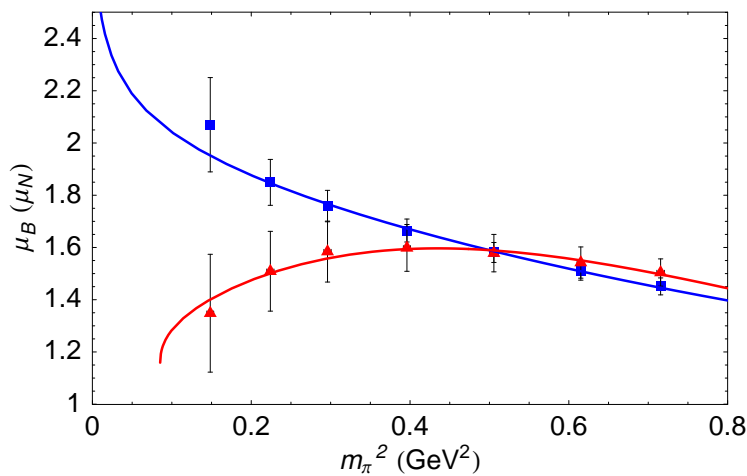}}
\vspace*{-1.0cm}
\caption{FLIC fermion simulation results for the magnetic moments of
  the proton ($\circ$) and $\Delta^+$ resonance ($\triangle$) in
  quenched QCD and the associated fits of FRR quenched \chiPT\
  to the lattice simulation results. }
\label{moments}
\end{center}
\vspace{-1.0cm}
\end{figure}

Figure \ref{moments} displays FLIC fermion simulation results for the
magnetic moments of the proton and $\Delta^+$ resonance in quenched QCD.
The curves illustrate the fits of FRR quenched \chiPT\ to the lattice
simulation results.  Here, the analytic terms of the chiral expansion
have been re-summed in a Pad\'e designed to reproduce the Dirac moment
mass dependence, $e \hbar / \,2 m$, at moderately large pion mass.

At large pion masses, the $\Delta$ moment is enhanced relative to the
proton moment in accord with earlier lattice QCD results
\cite{Leinweber:1990dv,Leinweber:1992hy} and model expectations.
However as the chiral regime is approached the nonanalytic behavior of
the quenched meson cloud is revealed, enhancing the proton and
suppressing the $\Delta^{+}$ in accord with the expectations of
quenched \chiPT.  The quenched artifacts of the $\Delta$ provide an
unmistakable signal for the onset of quenched chiral nonanalytic
behavior.

Supercomputing resources from the Australian Partnership for Advanced
Computing and research support from the Australian Research Council is
gratefully acknowledged.

\end{document}